\documentclass[english,aps,prX,showpacs,twocolumn]{revtex4-1}
\usepackage[T1]{fontenc}
\usepackage[latin9]{inputenc}
\usepackage{babel}
\usepackage{graphicx}
\usepackage{bm}
\usepackage{bbm}
\usepackage{array}
\usepackage{amsmath}
\usepackage{amssymb}
\usepackage{dcolumn}
\usepackage{epstopdf}
\usepackage{bbold}
\usepackage{tikz}
\usepackage[margin=15mm]{geometry}
\usetikzlibrary{shadings}

\usepackage{color}

\newcommand{\minline}[2]{$\mbox{\fontsize{#1}{10}\selectfont #2}$}

\begin{document}

\title{Relativistic  theory of spin relaxation mechanisms in the Landau-Lifshitz-Gilbert equation of spin dynamics}

\author{Ritwik Mondal}
\email[]{Ritwik.Mondal@physics.uu.se}
\author{Marco Berritta}
\author{Peter M. Oppeneer}
\affiliation{Department of Physics and Astronomy, Uppsala University, P.\,O.\ Box 516, Uppsala, SE-75120, Sweden}

\date{\today}

\begin{abstract}
Starting from the Dirac-Kohn-Sham equation we derive the relativistic equation of motion of spin angular momentum in a magnetic solid under an external electromagnetic field. This equation of motion can be rewritten in the form of the well-known Landau-Lifshitz-Gilbert equation for a harmonic external magnetic field, and leads to a more general magnetization dynamics equation for a general time-dependent magnetic field. In both cases with an electronic spin-relaxation term which stems from the spin-orbit interaction. We thus rigorously derive, from fundamental principles, a general expression for the anisotropic damping tensor which is shown to contain an isotropic Gilbert contribution as well as an anisotropic Ising-like and a chiral, Dzyaloshinskii-Moriya-like contribution. The expression for the spin relaxation tensor comprises furthermore both electronic interband and intraband transitions.  We also show that when the externally applied electromagnetic field possesses  spin angular momentum, this will lead to an optical spin torque exerted on the spin moment.
\end{abstract}

\pacs{75.78.-n, 76.20.+q, 71.15.Rf}
\maketitle

\section{Introduction}
In their seminal 1935-paper, L.\ D.\ Landau and E.\ M.\ Lifshitz proposed the equation of motion governing the dynamics of a continuum magnetization \cite{landau35}.  
Eighty  years after its original formulation, the  Landau-Lifshitz (LL) equation continues to play a fundamental role 
in the understanding of magnetization dynamics \cite{baryakhtar15} and forms the cornerstone of contemporary micromagnetic simulations 
(see, e.g.,\ Refs.\ \cite{brown63,kruzik06}).

Originally, the Landau-Lifshitz equation was derived on the basis of phenomenological considerations \cite{landau35}. It defines the time-evolution of a volume magnetization $\bm{M}(\bm{r},t)$ as
\begin{equation}
\frac{\partial \bm{M}}{\partial t} =- \gamma  \bm{M} \times \bm{H}_{\rm eff}   -\lambda \bm{M} \times [\bm{M} \times \bm{H}_{\rm eff}],
\label{LL}
\end{equation}
where $\gamma$ is the gyromagnetic ratio, $\bm{H}_{\rm eff}$ is the effective magnetic field, and $\lambda$ is an isotropic damping parameter. The first term 
describes the precession of the local magnetization $\bm{M} (\bm{r}, t)$ around the effective field $\bm{H}_{\rm eff}$.
The second term describes the magnetization relaxation such that the magnetization vector relaxes to the direction of the effective field. 
The damping term in the LL equation was reformulated by Gilbert \cite{gilbert56,gilbert04} to give the Landau-Lifshitz-Gilbert (LLG) equation,
\begin{equation}
\frac{\partial \bm{M}}{\partial t} =- \gamma  \bm{M} \times \bm{H}_{\rm eff}   + \alpha  \,  \bm{M} \times \frac{\partial \bm{M}}{\partial t },
\label{LLG}
\end{equation}
where $\alpha$ is the Gilbert damping constant. Note that both damping parameters $\alpha$ and $\lambda$ are here scalars, which corresponds to the assumption of an isotropic medium. Both LL and LLG equations preserve the length of the magnetization during the dynamics and are mathematically equivalent (see, e.g.\ \cite{lakshmanan11}).

 A number of explanations have been proposed for the microscopic origin of the spin relaxation in magnetic metals \cite{kambersky70,korenman72,baryakhtar84,garanin97,simanek03,tserkovnyak04,kambersky07,brataas08,garate09,hickey09,fahnle11}.
Already in their original work Landau and Lifshitz attributed the damping constant to  \textit{relativistic} effects \cite{landau35}.
More specific microscopic theories of spin relaxation  in ferromagnetic metals have been developed in the last decennia.
Kambersk{\'y} proposed the breathing Fermi surface model \cite{kambersky70} and the related torque-correlation model \cite{kambersky76,kambersky07}.
Brataas \textit{et al.}\ proposed a scattering theory formulation \cite{brataas08} of the Gilbert damping which is equivalent to a Kubo linear-response formulation.
A different form of the relaxation term caused by spatial dispersion of the exchange interaction---this in contrast to the isotropic medium assumption made in the LL equation---was proposed by Bar'yakhtar and co-workers \cite{baryakhtar84,baryakhtar13,baryakhtar13r}. 

More recently the debate on what the appropriate theory to describe damping would be has focused on first-principles electronic structure calculations and, in how far these could provide quantitative values of the Gilbert damping \cite{kunes02,steiauf05,gilmore07,starikov10,ebert11,sakuma12,barati14,thonig14,sakuma15}.
Recent \textit{ab initio} calculations of the Gilbert damping constant for transition-metal alloys predicted values that correspond to the experimental values within a range of a factor of two to three \cite{kunes02,steiauf05,gilmore07,ebert11,sakuma12,barati14}, with significant deviations however for the pure elemental ferromagnets. This indicates that there is still a need to improve the fundamental understanding of  the origin of spin-moment relaxation. Also, very recent publications have questioned the existing understanding of the Gilbert damping \cite{edwards16,li16}.

Here we develop a theoretical description of spin relaxation on the basis of the relativistic Density Functional Theory (DFT). To this end, we start from the relativistic Dirac-Kohn-Sham (DKS) equation that adequately describes the electronic states in a magnetic solid. From these we derive the general equation of motion for spin angular momentum, which adopts the form of the LLG equation.  Within this framework we obtain explicit expressions for the tensorial form of the Gilbert damping term, which we find to contain an isotropic Gilbert-like contribution and anisotropic Ising-like and chiral Dzyaloshinskii-Moryia-like contributions. 
Our derivation follows similar steps as a previous derivation by Hickey and Moodera \cite{hickey09}, however, as discussed below, it includes previously missing terms and thus leads to different expressions for the spin relaxation.


\section{The relativistic Dirac Hamiltonian}

\label{sec:Dirac}
As mentioned before, relativistic effects such as the spin-orbit interaction are at the heart of spin angular momentum dissipation in solids.  To examine how these fundamental physical interactions lead to magnetization damping we choose therefore to start from the most general relativistic Hamiltonian, the DKS Hamiltonian. This Hamiltonian describes the one-electron quantum state in an effective spin-polarized field due to other electrons and nuclei in the solid, in addition to externally applied fields. For spin-polarized electrons in a magnetic material the DKS Hamiltonian is given as \cite{macdonald79,eschrig99,greiner00}
\begin{eqnarray}
\mathcal{H}_{\rm D}&=&c\,\bm{\underline{\alpha}}\cdot\left(\bm{p}-e\bm{A}\right)+
\left(\underline{\beta}-\mathbb{\underline{1}}\right)mc^{2}+ 
V \mathbb{\underline{1}} + e\Phi \mathbb{\underline{1}} \nonumber \\
& &- \mu_B \underline{\beta} \, \bm{\underline{\Sigma}}\cdot\bm{B}^{\rm xc}.
\label{eq:DKSH_xc}
\end{eqnarray} 
Here $V$ is the unpolarized Kohn-Sham selfconsistent potential, 
$\bm{B}^{\rm xc}$ is the spin-polarized part of the exchange-correlation potential in the material, $\bm{A} = \bm{A}(\bm{r},t)$ is the vector potential of an externally applied electromagnetic field, $e\Phi (\bm{r},t)$ is the scalar potential of this field, $\bm{p}=-i\hbar\bm{\nabla}$,  and 
$\mu_B$ is $ \frac{e \hbar}{2m}$, the Bohr magneton.
$\underline{\mathbb{1}}$ is the $4\times4$ identity matrix and  $\underline{\bm{\alpha}}$, ${\underline{\beta}}$, and $\underline{\bm{\Sigma}}$
are the well-known Dirac matrices in Dirac bi-spinor space, which contain the Pauli spin matrices $\bm{\sigma}$ and 
the $2\times2$ identity matrix.
%
At this point, it is important
 to observe that there are two fundamentally different fields present in the DKS Hamiltonian. There are the Maxwell fields, that is, (implicitly) the  external magnetic induction
$\bm{B} (\bm{r}, t) =\bm{\nabla}\times\bm{A} (\bm{r},t) $ as well as the external electric field, $\bm{E} (\bm{r}, t) =-\frac{\partial \bm{A} (\bm{r}, t)}{\partial 
t} - \bm{\nabla}\Phi$. The strongest field in a magnetic material is however the exchange field, which stems from the Pauli exclusion principle. The exchange field $\bm{B}^{\rm xc}$ is fundamentally
different from the standard magnetic induction, as it obviously acts only on the spin degree of freedom (see, e.g., \cite{eschrig99}) and does not couple to the orbital 
angular momentum. Also, it doesn't fulfill the Maxwell equations as the 
auxiliary electromagnetic field (e.g., $\bm{\nabla} \cdot \bm{B} =0$) and it cannot be included as a vector potential $\bm{A}^{\rm xc}$ in the linear momentum, i.e.\ $\bm{p}-e\bm{A}^{\rm xc}$, but instead needs to be treated as a separate term in Eq.\ (\ref{eq:DKSH_xc}).

Next, we want to investigate the relativistic spin evolution of spin-polarized electrons in a magnetic solid. To achieve this we need the positive energy, that is, the electron solutions that are given by the large component of the Dirac bi-spinor.
To arrive at an elucidating formulation in terms of the spin operator we 
employ the
 Foldy-Wouthuysen transformation approach \cite{foldy50,greiner00} on the DKS equation for the case where an exchange  field $\bm{B}^{\rm xc}$ is explicitly present
 (for details, see Ref.\ \cite{mondal15}). Doing so, one obtains a Hamiltonian for the electron solutions only, which we expand in orders of  \minline{9}{$1/c^2$} to select the largest relativistic contributions. 
This leads to a semi-relativistic, extended Pauli Hamiltonian (see Ref.\ \cite{mondal15}),
\begin{widetext}
\begin{eqnarray}
\label{FW_hamiltonian}
\mathcal{H}_{\rm EP}&=&\frac{\left(\bm{p}-e\bm{A}\right)^{2}}{2m}+V  -  \mu_B \,\bm{\sigma}\cdot \bm{B} - \mu_B \,\bm{\sigma}\cdot \bm{B}^{\rm xc}_{\rm eff} +
e\Phi - \frac{\left(\bm{p}-e\bm{A}\right)^{4}}{8m^{3}c^{2}}-\frac{1}{8m^{2}c^{2}}\left(p^{2}V\right)-
\frac{e\hbar^{2}}{8m^{2}c^{2}}\bm{\nabla}\cdot\bm{E}\nonumber\\
&& +\frac{i}{4m^{2}c^{2}}\bm{\sigma}\cdot\left(\bm{p}V\right)\times\left(\bm{p}-e\bm{A}\right)-\frac{e\hbar}{8m^{2}c^{2}}\bm{\sigma}\cdot\left\{ \bm{E}\times\left(\bm{p}-e\bm{A}\right)-\left(\bm{p}-e\bm{A}\right)\times\bm{E}\right\}\nonumber\\
&& +\frac{i \mu_B}{4 m^{2}c^{2}}[(\bm{p}\times\bm{B}^{\rm xc})\cdot\left(\bm{p}-e\bm{A}\right)] .
\end{eqnarray}
 Except from the last term in Eq.\ (\ref{FW_hamiltonian}), all the appearing relativistic corrections involving the exchange interaction can be added together giving an \textit{effective} exchange field \cite{note1}, 
 \begin{eqnarray}
\bm{B}^{\rm xc}_{\rm eff}  & = & \bm{B}^{\rm xc}  - \frac{1}{8m^{2}c^{2}}\Big\{ \! \left[p^{2}\bm{B}^{\rm xc}\right]+2(\bm{p}\bm{B}^{\rm xc})\! \cdot \! \left(\bm{p}-e\bm{A}\right)+2(\bm{p}\cdot\bm{B}^{\rm xc})\!\left(\bm{p}-e\bm{A}\right)+4[\bm{B}^{\rm xc}\! \cdot \! \left(\bm{p}-e\bm{A}\right)] \!\left(\bm{p}-e\bm{A}\right)\Big\}
\nonumber \\
&\equiv & \bm{B}^{\rm xc} +  \bm{B}^{\rm xc}_{\rm corr}.
\label{Bxc-eff}
\end{eqnarray}
\end{widetext}
 The Hamiltonian   $\mathcal{H}_{\rm EP}$ exactly includes all spin-dependent relativistic terms (of the order of \minline{9}{$1/c^2$})  and all the terms involving $\bm{B}^{\rm xc}$ and the external electromagnetic fields. We emphasize that for our purpose of unveiling the relativistic mechanisms of spin dissipation it is obviously not sufficient to work with the conventional Pauli Hamiltonian,  which only consists of the five first terms in the nonrelativistic limit. The correct form of all  relativistic terms can solely be obtained when one starts from the DKS equation with exchange field. We remark that in a previous study Hickey and Moodera \cite{hickey09} used a Pauli Hamiltonian different from the above one, without exchange field and without crystal potential and thus without the intrinsic spin-orbit interaction [the first term in the second line of Eq.\ (\ref{FW_hamiltonian})].

The meaning of the terms in Hamiltonian (\ref{FW_hamiltonian}) 
can be readily understood, see Ref.\ \cite{mondal15} for details. 
The fourth term on the right is a Zeeman-like term due the presence of the relativistically corrected exchange field, which acts as an effective mean field. 
The ninth term is the one, which in a central potential $V$, gives rise to the conventional form of the spin-orbit coupling.  The tenth term is a kind of spin-orbit interaction but due to the external fields. The very last term is a relativistic correction which depends on the $\bm{B}^{\rm xc}$ field but is independent of the spin. 
As we will see in the following, the terms that are responsible for spin relaxation are the relativistic terms that involve a direct coupling of the spin operator with either the exchange field $\bm{B}^{\rm xc}$ or one of the externally applied fields ($\bm{E}$ or $\bm{A}$).

\section{Spin equation of motion}
The spin angular momentum operator is given by $\bm{S}=(\hbar/2)\bm{\sigma}$. To obtain an equation of motion for the spin operator we have to evaluate the commutator $[\bm{S},\mathcal{H}_{\rm EP}(t)]$. It is obvious from the expression of $\mathcal{H}_{\rm EP}$ that only the terms which are explicitly spin dependent will contribute as otherwise the commutator vanishes. We can thus extract from 
$\mathcal{H}_{\rm EP}$ the spin Hamiltonian
\begin{eqnarray}
\label{spin}
\mathcal{H}^{\bm{S}}(t) 
 &=& \mathcal{H}^0+\mathcal{H}_{\rm soc}^{\rm int} +\mathcal{H}_{\rm soc}^{\rm ext} 
\end{eqnarray}
where the Zeeman-like fields are added up to an effective magnetic induction,
\begin{eqnarray}
	\mathcal{H}^0 =-\frac{e}{m}\bm{S}\cdot\left(\bm{B}+\bm{B}^{\rm xc} + \bm{B}^{\rm xc}_{\rm corr}\right) \equiv -\frac{e}{m}\bm{S}\cdot\bm{B}_{\rm eff} .
\end{eqnarray}
The part $\mathcal{H}^0$ contains the main nonrelativistic contribution, all other terms in the spin Hamiltonian $\mathcal{H}^{\bm{S}}$ are of relativistic origin.
The intrinsic spin-orbit coupling is given by the Hamiltonian
\begin{eqnarray}
\label{Hintsoc}
	\mathcal{H}_{\rm soc}^{\rm int}= \frac{i}{2\hbar m^2c^2}\bm{S}\cdot(\bm{p}V)\times\left(\bm{p}-e\bm{A}\right) .
\end{eqnarray}
%
  The crystal potential stems from the nuclei-electron and electron-electron interactions and thus should have translational symmetry. Consequently, also the intrinsic spin-orbit Hamiltonian has translational symmetry \cite{cinal97}. If the position of any $j$-th nucleus is $\bm{R}_j$, the electron position is $\bm{r}$, and the electron position with respect to the nucleus is represented by $\bm{r}_{j}$, then the crystal potential can be represented by a sum of atom-centered potentials. Making now in addition the central potential approximation (no angular dependence) for each of the atom-centered potentials, 
the potential can be written as $V(r_{j}) = V(\vert \bm{r} - \bm{R}_{j} \vert )$. The translational symmetry is realized by the fact that $\bm{r}_{j} = \bm{r} - \bm{R}_{j}$. With the definition of spin-orbit interaction strength $\xi(r_j) = \frac{1}{2m^2c^2}\frac{1}{r}{dV(r_j)}/{dr} $, 
and the Coulomb gauge, $\bm{\nabla} \cdot \bm{A} = 0$, for homogeneous magnetic fields, i.e., $\bm{A}=(\bm{B}\times\bm{r})/2$, this Hamiltonian can further be written as
\begin{eqnarray}
\label{socint}
\!\!\!	\mathcal{H}_{\rm soc}^{\rm int} 
	&=& \frac{1}{2m^2c^2}\frac{1}{r}\frac{dV}{dr}\bm{S}\cdot\bm{L}-\frac{er}{4m^2c^2}\frac{dV}{dr}\bm{S}\cdot\bm{B}\nonumber\\
	&& +\frac{e}{4m^2c^2}\frac{1}{r}\frac{dV}{dr}\left(\bm{S}\cdot\bm{r}\right)\left(\bm{r}\cdot\bm{B}\right)\nonumber\\
	&=& \! \sum_j \xi(r_j)\!\left[ \bm{S}\cdot\bm{L}-\frac{e }{2}\Big( r^2 \bm{S}\cdot\bm{B} -\left(\bm{S}\cdot\bm{r}\right)\left(\bm{r}\cdot\bm{B}\right) \Big)\! \right]\!.\,\,\,\,\,\,
\end{eqnarray}
We note, first, that the full spin-orbit Hamiltonian, $\mathcal{H}_{\rm soc}^{\rm int} +\mathcal{H}_{\rm soc}^{\rm ext}$,  is gauge invariant \cite{mondal15b}, but for deriving expressions we need to make a choice. The Coulomb gauge is a suitable choice here,  yet it can be used exactly only when a slowly varying and homogeneous magnetic field is present. This gauge further implies that only the transversal parts of  $\bm{E}$ and of $\bm{A}$ are retained, the latter being gauge invariant.
Doing so, we have thus recovered the ``usual'' spin-orbit coupling term and other ultra-relativistic terms.

The external spin-orbit coupling Hamiltonian is given by
\begin{eqnarray*}
	\mathcal{H}_{\rm soc}^{\rm ext}=-\frac{e}{4m^2c^2}\bm{S}\cdot\left\{\bm{E}\times\left(\bm{p}-e\bm{A}\right)-\left(\bm{p}-e\bm{A}\right)\times\bm{E}\right\} ,
\end{eqnarray*}
 which has a similar form as $\mathcal{H}_{\rm soc}^{\rm int}$ [Eq.\ (\ref{Hintsoc})], but contains the external Maxwell fields instead.
Making use of Maxwell's equation $ \bm{\nabla} \times \bm{E} = - \partial \bm{B} / \partial t$, this Hamiltonian can be rewritten as
\begin{eqnarray}
\label{socext}
	\mathcal{H}_{\rm soc}^{\rm ext} &=& -\frac{e}{2m^2c^2}\bm{S}\cdot\left(\bm{E}\times\bm{p}\right)+\frac{ie\hbar}{4m^2c^2}\bm{S}\cdot \frac{\partial \bm{B}}{ \partial t} \nonumber\\
	&& +\frac{e^2}{2m^2c^2}\bm{S}\cdot\left(\bm{E}\times\bm{A}\right) .
\end{eqnarray}

The last term in the  Hamiltonian $\mathcal{H}_{\rm soc}^{\rm ext}$ describes the interaction of the photon spin angular momentum density, $\bm{j}_s=\epsilon_0(\bm{E}\times\bm{A})$ \cite{allen03}, with the electron spins \cite{mondal15b,bauke14}. A related interaction energy due to a coupling of the angular momentum density of the electromagnetic field with the magnetic moment  was proposed recently on phenomenological grounds \cite{bellaiche13}.
{The relativistic light-spin interaction  in the Hamiltonian (\ref{socext}) adopts thus the form
\begin{eqnarray}
	\mathcal{H}_{\rm light-spin}^{\rm ext}=\frac{e^2}{2m^2c^2\epsilon_0}\bm{S}\cdot\bm{j}_s .
\end{eqnarray}
This term, being second order in the external fields can become important in the strong field regime. As we focus in first instance on the damping, we will not consider it in the derivation of the spin damping, but we come back to it later on.

Now we have the necessary parts of the spin Hamiltonian and we are ready to calculate the spin dynamics equations.
According to the definition of magnetization, this quantity is given by the expectation value of spin angular momentum \cite{white07} 
\begin{equation}
\label{magn}
{\bm{M}}=\sum_{j}\frac{g\mu_B}{\mathcal{V}}\textrm{Tr} \left\{\rho{\bm{S}}^{j}\right\},
\end{equation}
where $\mathcal{V}$ is a suitably chosen volume element. 
The summation is taken over all the electrons $j$ and the definition of the density matrix is ${\rho}=\sum_i p_i |\psi_i\rangle\langle\psi_i|$, where the set of wave functions $|\psi_i\rangle$ are in a mixed state and $p_i$ are the occupation numbers. As is customary in spin dynamics models \cite{kambersky76,tserkovnyak04,simanek03,brataas08,kambersky07,hankiewicz08,garate09,hickey09,fahnle11,baryakhtar13,gilmore07,ebert11}
 the contribution of the orbital angular momentum to the total magnetization has been neglected 
because it is quenched for the common transition metals  (e.g., Fe, Ni, Co etc.).
The equation of motion of the magnetization is obtained by taking the time derivative on both sides of Eq.\ (\ref{magn}), and using that $\partial \rho/ \partial t=0$ for quasiadiabatic processes \cite{ho04},  which gives
\begin{equation}
\frac{\partial \bm{M}}{\partial t}=\frac{g\mu_B}{\mathcal{V}}\frac{1}{i\hbar}\sum_j\textrm{Tr}\left\{\rho[\bm{S}^j,\mathcal{H}^{\bm{S}}(t)]\right\}\, .
\label{eq:motion}
\end{equation}
To obtain the magnetization dynamics we substitute the spin Hamiltonian $\mathcal{H}^{\bm{S}}(t) 
 = \mathcal{H}^0+\mathcal{H}_{\rm soc}^{\rm int} +\mathcal{H}_{\rm soc}^{\rm ext}$
 in the right-hand side of Eq.\ (\ref{eq:motion}) and work out the trace term-by-term.
 
 Before presenting the result we consider briefly the approximations made in the derivation. Notably, Eq.\ (\ref{eq:motion}) is valid for local processes and will hence provide a \textit{local} damping mechanism. However, it is known that \textit{nonlocal} contributions to the damping exist (see, e.g., \cite{walowski08,nembach13,weindler14}) that can be caused by spin transport from one region to another \cite{mizukami02,tserkovnyak02,zhang09}. Such effects can be treated using the continuity equation, $\partial \rho / \partial t + \bm{\nabla} \cdot \bm{J} = 0$, with $\bm{J}$ the current operator, leading to an additional spin current term (see, e.g., \cite{zhang04,zhang09}). 
 A further remark due at this point concerns the time dependence of the exchange field. In line with the above,  we adopt the adiabatic approximation that is valid for systems not too far from the ground state \cite{marques04}.

Working out the commutator, we find that the first order dynamical equation of motion is given by the mostly nonrelativistic part in the spin Hamiltonian, $\mathcal{H}^0$. Using the commutation relations for spin angular momentum, $[S_{j},S_{k}]=i\hbar\epsilon_{jkl}S_{l}$, the first order equation of motion becomes
\begin{eqnarray}
\label{dynamics1}
\frac{\partial \bm{M}}{\partial t}\Big|^0= - \gamma\bm{M}\times\bm{B}_{\rm eff},
\end{eqnarray}
 where $\gamma= {g|e|}/{2m}$ is the gyromagnetic ratio and $g\approx 2$ for spin degrees of freedom. Using $\bm{B} =\mu_0( \bm{H} +  \bm{M})$, the right-hand term can be rewritten in the conventional form as $-\gamma_0 \bm{M}\times\bm{H}_{\rm eff}$, where $\gamma_0=\mu_0\gamma$. This equation provides the common understanding of the Larmor precessional motion of magnetization around an effective magnetic field, with a distinction that there is a relativistic correction $ \bm{B}^{\rm xc}_{\rm corr}$ to this field that has not been noted before.

Next we treat the relativistic spin-orbit effects in the magnetization dynamics. As we will see, these are the ones  that lead to local damping, i.e., the spin relaxation mechanisms in a magnetic solid are of relativistic origin \cite{landau35,korenman72}. First, we focus on the relativistic intrinsic spin-orbit coupling Hamiltonian $\mathcal{H}_{\rm soc}^{\rm int}$ in Eq.\ (\ref{socint}). Due to the quenching of the orbital angular momentum, the first term vanishes. The dynamics due to the remaining two terms in the Hamiltonian is calculated as
\begin{eqnarray}
\label{dynamics2}
	\frac{\partial \bm{M}}{\partial t}\Big|_{\rm soc}^{\rm int} \!\! &=& \frac{e}{4m^2c^2}\Big{\langle} r\frac{dV}{dr}\Big{\rangle}\bm{M}\times\bm{B}\nonumber\\
	&& ~~~~ -\frac{e}{4m^2c^2} \bm{M}\times\Big{\langle}\bm{r}\,\,\frac{1}{r}\,\frac{dV}{dr}\,\left(\bm{r}\cdot\bm{B}\right)\big\rangle \nonumber \\
	 &=&  \frac{e}{2}\sum_{j} \Big[\langle \xi(r_j) r^2 \rangle\bm{M}\times\bm{B}   \nonumber \\
	 && ~~~~~ -\bm{M}\times \langle \xi(r_j)\bm{r}\,\left(\bm{r}\cdot\bm{B}\right)\rangle \Big]  .\,\,\,\,\,\,\,\,
\end{eqnarray}
The first term in the dynamics of Eq.\ (\ref{dynamics2}) can be seen as a further relativistic correction to the magnetization precession. The second term has a form similar to the first term, but with opposite sign. The terms can be combined, but they do not contribute to any relaxation processes as they do not contain a time variation of the magnetic induction.

Next we consider the dynamics related to $\mathcal{H}_{\rm soc}^{\rm ext}$.  We will see below that
it is mainly the relativistic extrinsic spin-orbit coupling, i.e., the first two terms of Eq.\ (\ref{socext}), which give rise to dominant local spin relaxation mechanisms in magnetic solids. In addition, we observe here that these correspond to the transverse spin relaxation. 
We consider here the long wavelength approximation, where the wavelength of the field is much larger than the size of the system. 
In other words the GHz/THz electromagnetic field inside the ferromagnetic film is assumed uniform throughout the film as long as the film thickness is sufficiently  small. We can thus use the Coulomb gauge, i.e., $\bm{A}=(\bm{B}\times\bm{r})/{2}$.
This gauge allows us to obtain the explicit time dependence of the Hamiltonian.
The transverse electric field in the Hamiltonian is then 
written as 
$\bm{E}= \frac{1}{2}\left( \bm{r} \times \partial \bm{B} / \partial t \right)$.
Employing the gauge, the first two terms in Eq.\ (\ref{socext}) can be re-written in an explicit, time-dependent form:
\begin{eqnarray}
\label{socext_new}
\mathcal{H}_{\rm soc}^{\rm ext} 
&=& \frac{ie\hbar}{4m^{2}c^{2}}\bm{S}\cdot \frac{\partial \bm{B}}{\partial t} \left(1-\frac{(\bm{r}\cdot\bm{p})}{i\hbar}\right)\nonumber\\
&& +\frac{e}{4m^2c^2}\left(\bm{S}\cdot\bm{r}\right)\left(\frac{\partial \bm{B}}{\partial t} \cdot\bm{p} \right).
\end{eqnarray}

 At this point it is needed to inspect the hermiticity of the Hamiltonian. It can be shown that the total spin-orbit Hamiltonian  in Eq.\ (\ref{socext_new}) is hermitian (see Appendix A), however, for the individual terms it is different.  Writing down the Hamiltonian in component form with the usual summation convention, we obtain
\begin{eqnarray}
\mathcal{H}_{\rm soc}^{\rm ext}&=&  \underbrace{\frac{ie\hbar}{4m^{2}c^{2}}S_i \frac{\partial B_i}{\partial t}}_{\rm  anti-hermitian}-\underbrace{\frac{e}{4m^{2}c^{2}}\sum_{i\neq j}S_i \frac{\partial B_i}{\partial t} r_jp_j}_{ \rm non-hermitian}\nonumber\\
&+&\underbrace{\frac{e}{4m^{2}c^{2}}\sum_{i \neq j} S_ir_i \frac{\partial B_j}{\partial t} p_j}_{ \rm hermitian} .
\end{eqnarray}
Previously, Hickey and Moodera considered the effect of the spin-orbit Hamiltonian on damping, but only obtained the first two terms in Eq.\ (\ref{socext}) \cite{hickey09}. They proposed then only the anti-hermitian part of the Hamiltonian as an intrinsic source of Gilbert damping \cite{hickey09}. Anti-hermitian Hamiltonians understandably are always dissipative \cite{Widom09,galda15}. Consequently, their choice of taking the anti-hermitian term only was criticized, given that the full spin-orbit Hamiltonian should be hermitian and that it therefore should not exhibit dissipation \cite{Widom09}.

 In our case the total spin-orbit  Hamiltonian (\ref{socext_new}) is manifestly hermitian, yet we will show below that it does give rise to spin moment damping. 
The point is, that even when the full Hamiltonian is hermitian, it only has this property when one considers the dynamics of the full system. It is however customary in spin moment dynamics \cite{kambersky76,tserkovnyak04,simanek03,brataas08,kambersky07,hankiewicz08,garate09,hickey09,fahnle11,baryakhtar13,gilmore07,ebert11} to integrate out the orbital degree of freedom and other magnetic degrees of freedom (as background fluctuations of the system) thus restricting the focus on the single spin moment dynamics. In the thereby restricted Hilbert space the hermiticity is lost and hence the whole Hamiltonian can contribute to the damping.


Calculating now the commutation relation $[\bm{S},\mathcal{H}_{\rm soc}^{\rm ext}]$ and taking the summation of the trace over all electrons, the spin moment dynamics adopts the form
\begin{eqnarray}
\label{dynamics3}
	\frac{\partial \bm{M}}{\partial t}\Big|_{\rm soc}^{\rm ext} &=& -\frac{ie\hbar}{4m^{2}c^{2}}\bm{M}\times \frac{\partial \bm{B}}{\partial t}\left(1-\frac{\big{\langle}\bm{r}\cdot\bm{p}\big{\rangle}}{i\hbar}\right)\nonumber\\
	&& -\frac{e}{4m^2c^2}\bm{M}\times\Big{\langle}\bm{r}\left( \frac{\partial \bm{B}}{\partial t} \cdot \bm{p} \right)\Big\rangle.
\end{eqnarray}
A rewriting of these terms is required to elucidate further the spin relaxation.

\section{The damping equations}
\label{damp-equations}

To obtain explicit expressions for the damping terms, we employ the
 general relation between magnetic induction $\bm{B}$, magnetization $\bm{M}$, and magnetic field $\bm{H}$, given as $\bm{B} = \mu_0 (\bm{M} + \bm{H})$. We take the time derivative on both sides,
\begin{eqnarray}
\label{variation}
	\frac{\partial \bm{B}}{\partial t} & = & \mu_0 \left[ \frac{\partial \bm{M}}{\partial t}+ \frac{\partial \bm{H}}{\partial t}\right].
\end{eqnarray}  
This relation is generally valid, also for the stationary case, even though the magnetization $\bm{M}(t)$ and magnetic field $\bm{H}(t)$ are time dependent. 
At this point it is instructive to consider what kinds of magnetic fields $\bm{H}(t)$ can occur. The simplest case is when 
at some time $t_0$ only a \textit{static} field $\bm{H}_0$ is present, then obviously only the first term in Eq.\ (\ref{variation}) contributes. 
If the field $\bm{H}(t)$ is explicitly time dependent, we can distinguish to cases: an ac driven, periodic magnetic field, as is commonly used in measurements, or a more general field, for example a magnetic field pulse.
In the latter case, one could proceed to derive the spin dynamics by keeping explicitly the term  $\frac{\partial \bm{H}}{\partial t}$. As a result, one obtains a LLG-like equation, where however the magnetic field couples into the damping term. The thus-obtained modified LLG equation is given and analyzed further below, in Sect.\ \ref{Discussion}.

In the former case, the effect of the magnetic response becomes apparent when an ac magnetic field is applied. For ferromagnetic materials, where there is a net magnetization present even in the absence of the applied field, the magnetic susceptibility can be introduced by the definition: $\chi = {\partial\bm{M}}/{\partial \bm{H}}$. Using a chain rule for the time derivative, $\frac{\partial \bm{H}}{\partial t} = \frac{\partial \bm{H}}{\partial \bm{M}} \frac{\partial \bm{M}}{\partial t}$, 
Eq.\ (\ref{variation}) can be written as
\begin{eqnarray}
	\frac{\partial\bm{B}}{\partial t}=\mu_0 \left(\mathbb{1} + {\chi}^{-1} \right) \cdot\frac{\partial\bm{M}}{\partial t}\,,
	\label{dBdt}
\end{eqnarray}   
where $\mathbb{1}$ is the $3 \times 3$ identity matrix. This relation has been used in the ensuing magnetization dynamics.

 Substituting Eq.\ (\ref{dBdt}) in the first term of Eq.\ (\ref{dynamics3}), we obtain 
\begin{eqnarray}
\frac{\partial \bm{M}^{(1)}}{\partial t}\Big|_{\rm soc}^{\rm ext} \!\! =-\frac{ie\hbar\mu_0}{4m^{2}c^{2}}\bm{M} \! \times \! \left[(\mathbb{1}+{\chi}^{-1})\! \cdot \! \frac{\partial \bm{M}}{\partial t} \right] \!\! \left(1-\frac{\langle\bm{r}\cdot\bm{p}\rangle}{i\hbar} \right) \! .\nonumber\\
\label{soc-gilbert}
\end{eqnarray}
This term can already be recognized to have the form of the Gilbert damping, $\bm{M}\times\left[{\alpha} \cdot \frac{\partial \bm{M}}{\partial t}\right]$, yet with a tensorial damping constant.

 For the full damping we have to combine with the second term in Eq.\ (\ref{dynamics3}), which is rewritten as
\begin{eqnarray}
\frac{\partial \bm{M}^{(2)}}{\partial t}\Big|_{\rm soc}^{\rm ext} \!\!
&=& - \frac{e\mu_0}{4m^2c^2}\bm{M} \! \times \! \Big{\langle}\bm{r}\left( \Big[(\mathbb{1}+\chi^{-1})\cdot\frac{\partial \bm{M}}{\partial t}\Big] \cdot \bm{p} \right) \! \Big\rangle  . \nonumber\\
\label{soc-gilb2}
\end{eqnarray}
 To join the terms we proceed with using vector components. Equation (\ref{soc-gilbert}) becomes
\begin{eqnarray}
	\frac{\partial \bm{M}^{(1)}}{\partial t}\Big|_{\rm soc}^{\rm ext} \!\! 
	=& -& \frac{e\mu_0}{4m^{2}c^{2}}\sum_{ijkln} M_k\left[(\mathbb{1}+{\chi}^{-1})_{ij} \frac{\partial M_j}{\partial t} \right]  \nonumber \\
	& & ~~~ \times \left(i\hbar - {\langle {r}_n {p}_n\rangle}  \right)\varepsilon_{kil}\hat{\bm{e}}_{l} ,
\end{eqnarray}
with $\varepsilon_{ijk}$ the Levi-Civita tensor and $\hat{\bm{e}}$ a unit vector. 
This term can be written as
\begin{eqnarray}
	\frac{\partial \bm{M}^{(1)}}{\partial t}\Big|_{\rm soc}^{\rm ext} \!\! 
	&=& \sum_{ijkl} M_k\, \frac{\partial M_j}{\partial t}\, \Omega_{ij}\,\varepsilon_{kil}\,\hat{\bm{e}}_{l},
\end{eqnarray}
with $\Omega_{ij} =-\frac{e\mu_0}{4m^{2}c^{2}}\sum_{n}\left(i\hbar-\langle {r}_n {p}_n\rangle \right)(\mathbb{1}+{\chi}^{-1})_{ij}$.
The second term (\ref{soc-gilb2}) can be written in a similar form, but with a tensor $\Delta_{ij} = -\frac{e\mu_0}{4m^2c^2}\sum_{n}\langle {r}_i {p}_n \rangle(\mathbb{1}+\chi^{-1})_{nj}$.
Combining these two terms gives the total damping term,
\begin{eqnarray}
\frac{\partial \bm{M}}{\partial t}\Big|_{\rm soc}^{\rm ext} \!\! =\sum_{ijkl}M_k \Big[\Omega_{ij}+\Delta_{ij}\Big]\frac{\partial M_j}{\partial t}\varepsilon_{kil}\hat{\bm{e}}_l ,
\end{eqnarray}
where it is convenient to define $\textrm{A}_{ij} \equiv \Omega_{ij}+\Delta_{ij}$,
\begin{eqnarray}
\! \! \! \! \! \! \textrm{A}_{ij} \! & = & -\frac{e\mu_0}{4m^{2}c^{2}}\sum_{n} \! \Big[i\hbar-\langle {r}_n {p}_n\rangle +\langle {r}_n {p}_i \rangle\Big] \! (\mathbb{1}+{\chi}^{-1})_{ij}\nonumber\\
& = &  -\frac{e\mu_0}{8m^{2}c^{2}}\sum_{n,k} \! \Big[  \langle {r}_i {p}_k + p_kr_i\rangle -\langle r_np_n + p_nr_n \rangle \delta _{ik} \Big]  \nonumber\\&& \! \! \! \! \! \! \! \qquad \qquad \qquad \qquad \qquad \qquad \qquad  \times \,\, \! (\mathbb{1}+{\chi}^{-1})_{kj}.
\label{Gilbert-tensor}
\end{eqnarray}
 Note that a summation over $i$ is not intended in the right-hand side expressions.
In vector form the spin-orbit damping term becomes
\begin{equation}
\frac{\partial \bm{M}}{\partial t}\Big|_{\rm soc}^{\rm ext} = \bm{M}\times\Big[\textrm{A}\cdot\frac{\partial \bm{M}}{\partial t}\Big].
\label{LLG-A}
\end{equation}
Summarizing our result, we observe that we have obtained a damping parameter $\textrm{A}_{ij}$ of Gilbert type that is however in its general form not a scalar but a tensor.
The tensorial character of the Gilbert damping was also concluded recently in other investigations \cite{gilmore10,garate09}. In this form it accounts for transversal spin relaxation that conserves the length of the magnetization, i.e., $\partial (\bm{M} \cdot \bm{M}) / \partial t =0$.

Every tensor can be decomposed in a symmetric and an anti-symmetric part. Hence, the damping tensor can  be decomposed 
 into a scalar ($\alpha$) multiplied by the unit matrix, a symmetric tensor ($\mathbb{I}$), and an anti-symmetric tensor ($\mathbb{A}$,  with $\mathbb{A}_{ij} = \frac{1}{2}\left(A_{ij}-A_{ji}\right)$). The latter tensor can in turn be expressed as $\mathbb{A}_{ij}=\varepsilon_{ijk}\bm{D}_k$ with $\bm{D}$ being a vector. Finally, the damping dynamics can then be written as
\begin{eqnarray}
\frac{\partial \bm{M}}{\partial t}\Big|_{\rm soc}^{\rm ext} \!\! &=& \alpha \,\bm{M}\times\frac{\partial \bm{M}}{\partial t} +
\bm{M}\times\Big[\mathbb{I}\cdot\frac{\partial \bm{M}}{\partial t}\Big] \nonumber \\
 && + \bm{M}\times\Big[\bm{D}\times\frac{\partial \bm{M}}{\partial t}\Big] .
\end{eqnarray}
The first term is the conventional Gilbert damping. It originates from the decomposition of the symmetric part of the tensor into an isotropic Heisenberg-like $(\alpha \mathbb{1})$ contribution as well as an anisotropic Ising-like ($\mathbb{I}$) contribution which leads to the second term. Along with that it is not surprising that the last term implies a Dzyaloshinskii-Moriya-like contribution. The anisotropic nature of the Gilbert damping has been noted before \cite{gilmore10,fahnle11}, but not the appearance of the Dzyaloshinskii-Moriya-like damping. This type of damping could be related to the chiral damping of magnetic domain walls that was reported recently \cite{jue16}.

For the case of a constant, scalar Gilbert damping parameter it is straightforward to transform the LLG equation to obtain the LL equation with the phenomenological damping term proposed by Landau and Lifshitz \cite{landau35}. However, this is no longer the case for tensorial Gilbert damping, for which the transformation is much more involved. The spin-dynamics equation in the Landau-Lifshitz form now becomes (see Appendix B)
	\begin{eqnarray}
\!\!	\! \! &&\left(\Psi^2\mathbb{1}+\mathbb{G}\right)\cdot\frac{\partial\bm{M}}{\partial t} = \nonumber \\
	&&-\gamma_0\Psi\bm{M}\times\bm{H}_{\rm eff}-\gamma_0\bm{M}\times \! \Big[ \!\! \left(\alpha\mathbb{1}+\mathbb{I}\right) \! \cdot \! (\bm{M}\times\bm{H}_{\rm eff})\Big],\,
	\label{LL-full}
\end{eqnarray}
where $\Psi=1+\bm{M}\cdot\bm{D}$ and the tensor $\mathbb{G}$ is defined through
\begin{eqnarray}
	\mathbb{G}&=& \alpha^2\,M^2\mathbb{1}-\Big[\left(\bm{M}\cdot\mathbb{I}\cdot\bm{M}\right)-\mathfrak{t} M^2\Big]\left(\alpha\mathbb{1}+\mathbb{I}\right)	\nonumber\\
		&-&\Big( \mathfrak{t} \bm{M}-\bm{M}\cdot\mathbb{I}\Big)\bm{M}\cdot\mathbb{I}-
	M^2\mathbb{I}^2+\bm{M}\Big(\bm{M}\cdot\mathbb{I}^2\Big) ,
\end{eqnarray}
with the trace,  $\mathfrak{t}={\rm Tr}\,(\mathbb{I})$.  In general the trace of such a matrix $\mathbb{I}$ is non-zero, however its value will depend on how the symmetric tensor $A_{ij}^{\rm sym} = \frac{1}{2} \left(A_{ij}+A_{ji}\right) = \mathbb{I}_{ij} + \alpha \delta _{ij}$ is decomposed. If the decomposition in Ising and Heisenberg parts is such that the isotropic part is chosen as $\alpha = \frac{1}{3} {\rm Tr}(A^{\rm sym}_{ij})$, 
then the trace of $\mathbb{I}$ will vanish, $\mathfrak{t} = 0$ .
Note that the term (\ref{dynamics2}) due to the intrinsic spin-orbit interaction has been left out, as it is expected to give only a small correction to the effective magnetic field.
The damping term thus adopts the form 
$-\gamma_0\bm{M}\times\left[\Lambda \cdot\left(\bm{M}\times\bm{H}_{\rm eff}\right)\right]$,
similar to the phenomenological damping considered by Landau and Lifshitz  \cite{landau35}, but with damping tensor $\Lambda $. A more general form of the LL damping as  a tensor was already considered much earlier (see, e.g.\ \cite{Akhiezer68}), and it is reflected also in our derivation. However, a distinction is that here the leading $\partial \bm{M} / {\partial t}$ term on the left-hand side in Eq.\ (\ref{LL-full}) is, in its general form, multiplied not with a scalar ($1 + \alpha^2 M^2$) but with a tensor which moreover depends on the direction of $\bm{M}$. 

 It is worth noting that in the absence of the Dzyaloshinskii-Moriya and anisotropic relaxation contributions, i.e., setting $\bm{D}=\mathbb{I} =0$ we retrieve the original LL and LLG equations with scalar damping parameters. The validity range of our derived equations of spin motion is thus larger than the originally proposed equations of motion. It should also be emphasized that the Dzyaloshinskii-Moriya-like contribution appears in the Gilbert damping, however, it does not appear in the damping term of the LL equation (\ref{LL-full}). Instead, it leads to the \textit{renormalization} of the standard dynamical terms in the LL equation as can be seen from the appearance of the quantity $\Psi$ in Eq.\ (\ref{LL-full}).
We lastly note that the here obtained relaxation terms 
do not allow a variation with respect to the coordinates i.e., they do not include effects of spatial dispersion.

\section{Discussion}
\label{Discussion}

\subsubsection{Analysis of the damping expression}

Equation (\ref{Gilbert-tensor}) for the Gilbert damping pertains to the relaxation of spin motion in the presence of spin-orbit interaction. This damping is of relativistic origin as is exemplified by its \minline{9}{$1/c^2$} dependence. The expression for the Gilbert tensor is different from that obtained previously \cite{hickey09}, where only the constant term $i\hbar$ in the square bracket was found. 
The new parts $\langle {r}_i {p}_j\rangle$ relate to how the electronic band energies $E_{\nu\bm{k}}$ of Bloch states $| \nu \bm{k} \rangle$ disperse with $k$-space direction. It can be 
rewritten as  (see Appendix \ref{ab-initio})
\begin{eqnarray}
\langle {r}_{i} {p}_{j}\rangle &=& -\frac{i\hbar}{2m}\sum_{\nu, \nu^{\prime} ,\bm{k} }\frac{f(E_{\nu \bm{k}}) -f (E_{\nu^{\prime}\bm{k}})}{ E_{\nu\bm{k}}-E_{\nu^{\prime}\bm{k}} } p_{\nu\nu^{\prime}}^{i}p_{\nu^{\prime}\nu}^{j},
\end{eqnarray}
where $\bm{p}_{\nu\nu'} \equiv \langle \nu \bm{k} | \bm{p} | \nu' \bm{k} \rangle$ and $f(E_{\nu \bm{k}})$ is the Fermi function. The sum contains interband and intraband contributions. The intraband (Fermi surface) contribution  $(\nu=\nu^{\prime})$  can be written as
\begin{eqnarray}
\langle {r}_{i}{p}_{j}\rangle = 
         - \frac{i m}{2\hbar}\sum_{\nu \bm{k}}\left(\frac{\partial f}{\partial E}\right)_{\! \! E_{\nu \bm{k}}} \! \! \!  \left(\frac{\partial E_{\nu \bm{k}}}{\partial k_i}\right)
         \left( \frac{\partial E_{\nu \bm{k}}}{\partial k_j} \right).
\label{eq:intraband}         
\end{eqnarray}
This expression has a similarity with other previously derived expressions, as e.g.\ the breathing Fermi surface model \cite{kambersky70,gilmore07}  that has been applied to metallic ferromagnets.
The expression for the  $\langle {r}_i {p}_j\rangle$ terms has furthermore a form similar to that for the conductivity tensor in linear-response theory \cite{oppeneer01}; it is in particular well-suited for \textit{ab initio} calculations. 
 We note further that the influence of electron interaction with quasiparticles can be introduced by replacing $E_{\nu \bm{k}} -E_{\nu' \bm{k}}$ by  $E_{\nu \bm{k}} -E_{\nu' \bm{k}} + i\delta$, where the small $\delta$ gives a finite relaxation time to the electronic states.

For numerical evaluation of the damping tensor the susceptibility tensor $\chi$ is furthermore needed, which is in general wavevector and frequency dependent,  $\chi (\bm{q}, \omega )$. 
Thus, also the Gilbert damping tensor is here a frequency and $q$-dependent quantity, in accordance with recent measurements \cite{li16}.
 Suitable expressions for $\chi$ have been considered previously in the context of Gilbert damping \cite{tserkovnyak04,hankiewicz08,garate09}.
 Linear-response formulations that express $\chi$ as a spin-spin correlation function include the Pauli and Van Vleck susceptibility  contributions \cite{ghosh2005}, and expressions for the orbital susceptibility have been derived as well \cite{deng2000}.
These expressions are fitting for \textit{ab initio} calculations of $\chi$ within a DFT framework. The spin-orbit interaction will have an additional influence on $\chi$, however, unlike the main Gilbert damping contribution which is proportional to the spin-orbit coupling, this will only be a higher order effect.

We can consequently distinguish here two origins for the damping: the first one is related to the terms $\langle {r}_i {p}_j \rangle$, which represent dissipation contributions into the  orbital degrees of freedom. The second nature is due to the magnetic susceptibility $\chi$ which represents losses through the magnetic structure of the material. Both effects are simultaneously present,  and nonzero, for metallic ferromagnets as well as insulators.

It is also important to mention that the damping tensor in the our derivation does not include spin-relaxation effects due to interaction of spin-polarized electrons with quasiparticles as magnon or phonons or scattering with defects. Longitudinal spin relaxation due to spin-flip processes caused by electron-phonon scattering have been recently calculated \textit{ab initio} for the transition-metal ferromagnets \cite{carva13,essert11,illg13}, and magnon spin-flip scattering has been considered as well \cite{haag14}. Spin angular momentum  transfer due to explicit coupling of the spins to the lattice has been treated in several models \cite{vittoria10,baral14,garanin15}. As mentioned above, although the spin-lattice dissipation channel is not encompassed in our derivation,
an approximate way to include its influence  has been introduced before, by a suitable spectral broadening of the Bloch electron energies (see, e.g., \cite{sinova04,gilmore07}).

Lastly,  we remark that in the present derivation we obtain only first-order time-derivatives of $\bm{M}(\bm{r},t)$. Second-order time-derivatives of $\bm{M}(\bm{r},t)$ have recently been related to moment of inertia of the magnetization \cite{bhattacharjee12}.

\subsubsection{Exchange field and nonlocal contributions}

Thus far we have not explicitly discussed the exchange interaction. The influence of the exchange field can be accounted for in various levels of approximation, for example, within the Heisenberg model or evaluated within time-dependent DFT \cite{capelle01,qian02}. In the former, a suitable simplification of the exchange interaction in a magnetic solid is to express it through the Heisenberg Hamiltonian $\mathcal{H}^{\rm xc} = - \sum_{\alpha > \beta} J_{\alpha \beta} \textbf{S}_{\alpha} \cdot \textbf{S}_{\beta}$, where the $J_{\alpha \beta}$ are exchange constants and $\textbf{S}_{\alpha}$ is the atomic spin on atom $\alpha$. Using this Hamiltonian to express the exchange field leads to Landau-Lifshitz-Gilbert equations of motion for the dynamics of atomic moments (see, e.g., \cite{nowak05,evans14,ellis15}). 

More general, the exchange field depends on the spatial position which implies that there can exist an influence of  spatial nonuniformity of the exchange field on the spin relaxation. An influence on the dynamics occurring due to magnetization inhomogeneity  ($\nabla^2 \bm{M}$) appearing in the effective field was already suggested by Landau and Lifshitz \cite{landau35}. Such a term is in fact needed to properly describe spin wave dispersions \cite{herring51}.  A nonlocal damping mechanism due to spatial dispersion of the exchange field was proposed by Bar'yakhtar on the basis of phenomenological considerations such as symmetry arguments and Onsager's relations \cite{baryakhtar84}.  This leads to a modified expression for the damping term in the Landau-Lifshitz-Bar'yakhtar equation which contains the derivative of the exchange field $\nabla^2 \bm{B}^{\rm xc} $ \cite{baryakhtar84,baryakhtar13}. The existence of such nonlocal damping term can be related to the continuity equation connecting the spin density and spin current; it is  important for obtaining the correct asymptotic behavior of spin wave damping at large wavevectors $k$ \cite{baryakhtar13} known for magnetic dielectrics, see \cite{Akhiezer68}. Such nonlocal damping is important, too,  for describing spin current flow in magnetic metallic heterostructures \cite{yastremsky14}. These nonlocal damping terms are furthermore related to the earlier proposed magnetization damping effects due to spin diffusion \cite{baryakhatar1989doaminwalls,galkina1993,zhang09,tserkovnyak09} that have been studied recently \cite{wang15}. As a consequence of the spin current flow the local length of the magnetization is not conserved.
In the present work such nonlocal terms are not included since we focus on the local dissipation and have thus omitted the spin current contribution of the continuity equation.
A future full treatment that takes into account both local and nonlocal spin dissipation mechanisms would permit to describe magnetization dynamics and spin transport on an equal footing in a broader range of inhomogeneous systems.

\subsubsection{General time-dependent magnetic fields}

When the driving magnetic field is not an ac  harmonic field the dependence of $\bm{M}(\bm{r}, t)$ on $\bm{H}(t)$ will induce a more complex dynamics. In this case it is possible to derive a closed expression for the spin dynamics by explicitly keeping the term $\frac{\partial H}{\partial t}$ in Eq.\ (\ref{variation}). 
A similar derivation as presented in Sect.\ \ref{damp-equations} for the ac driving field leads then to the following expression for the magnetization dynamics
\begin{eqnarray}
\label{generaldMdt}
\!	\! \! \! \frac{\partial\bm{M}}{\partial t} \! = -\gamma_0\bm{M}\times\bm{H}_{\rm eff}+\bm{M}\! \times \! \Big[ \bar{\textrm{A}} \! \cdot \! \Big( \frac{\partial\bm{M}}{\partial t} +   \frac{\partial \bm{H}}{\partial t} \Big) \Big] \, ,
\end{eqnarray}
where the damping tensor $\bar{\textrm{A}}$ is given by 
\begin{equation}
 \bar{\textrm{A}}_{ij} \! =  -\frac{e\mu_0}{8m^{2}c^{2}}\sum_{n} \! \Big[  \langle {r}_i {p}_j + p_j r_i\rangle -\langle r_np_n + p_nr_n \rangle \delta _{ij} \Big] .
\label{Abar}
\end{equation}
The time-dependent magnetic field thus leads to a new, modified spin dynamics equation which has, to our knowledge, not been derived before.
The time-derivate of $\bm{H}(t)$ introduces here an additional torque,  $\bm{M}\! \times \!  \frac{\partial \bm{H}}{\partial t}$. This field-derivative torque might offer new ways to achieve fast magnetization switching. Consider for example an initially steep magnetic field pulse that thereafter relaxes slowly back to its initial value. The derivative of such field will exert a large but shortly lasting torque on the magnetization, which could initiate switching. Irradiation of magnetic thin films with a  picosecond THz field pulse was recently shown to trigger ultrafast magnetization dynamics \cite{vicario2013}, and suitable shaping of the THz magnetic field pulse could hence offer a route to achieve switching on a picosecond time scale.

\subsubsection{The optical spin torque}

The interaction of the spin moment with the optical spin angular moment $\bm{j}_s$ is given by the Hamiltonian $\mathcal{H}_{\rm light-spin}^{\rm ext}$. We note that such relativisitic interaction is important for recent attempts to manipulate the magnetization in a material using optical angular momentum, i.e.,\ helicity of the laser field \cite{tesarova13,lambert14,mondal15b}. This interaction leads to spin dynamics of the form
\begin{eqnarray}
	\frac{\partial \bm{M}}{\partial t}\Big\vert_{\rm light-spin}^{\rm ext}
		&=& -\frac{e^2}{2m^2c^2\epsilon_0}\,\bm{M}\times\bm{j}_s ,
\label{opt-spintorque}
\end{eqnarray} 
where $\bm{M}\times\bm{j}_s $ is the optical spin torque exerted by the optical angular moment on the spin moment. This equation expresses that the spin moment in a material  can be manipulated by acting on it with the optical spin angular moment of an external electromagnetic field in the strong field regime.

\section{Conclusions}
On the basis of the relativistic Dirac-Kohn-Sham equation we have derived the spin Hamiltonian to describe adequately the dynamics of electron spins in a solid, taking into account all the possible spin-related relativistic effects up to the order \minline{9}{$1/c^2$} and the exchange field and external electromagnetic fields. From this manifestly hermitian spin Hamiltonian we have calculated the spin equation of motion which adopts the form of the Landau-Lifshitz-Gilbert equation  for applied harmonic  fields. For universal time-dependent external magnetic fields we obtain a more general dynamics equation which involves the field-derivative torque. Our derivation does notably not rely on phenomenological assumptions but provides a rigorous treatment on the basis of fundamental principles, specifically,  Dirac theory with all relevant fields included. 

We have shown the existence of a relativistic correction to the precessional motion in the obtained LLG equation  and have derived an expression for the spin relaxation terms of relativistic origin.  One of the most prominent results of the presented article  is the derived expression for the tensorial Gilbert damping, which has been  shown to contain an isotropic Gilbert contribution, an anisotropic Ising-like contribution, and a chiral, Dzyaloshinskii-Moriya-like contribution. 
 Transforming the LLG equation to the Landau-Lifshitz equation of motion, we showed that the LLG equation with anisotropic tensorial Gilbert damping cannot trivially be written as a LL equation with an anisotropic LL damping term, but an additional matrix appears in front of the $\partial \bm{M} / \partial t$ term.  The Dzyaloshinskii-Moriya-like contribution serves as a renormalization factor to the common LL dynamical terms. The obtained} expression for the Gilbert damping tensor  in the case of a periodic driving field depends on the spin-spin susceptibility response function along with a term representing the electronic spin damping due to dissipation into the orbital degrees of freedom. As there exist an on-going discussion on what the fundamental origin of the Gilbert damping is and how it can accurately be evaluated from first-principles calculations \cite{barati14,sakuma15,edwards16,li16}, we point out that the two components of the derived damping expression (spin-spin and current-current response functions) are suitable for future \textit{ab initio} calculations within the density functional formalism. \\


\begin{acknowledgments}

 We thank B.\ A.\ Ivanov, P.\ Maldonado, A.\ Aperis, K.\ Carva, and H.\ Nembach  for helpful discussions.  We also thank the anonymous reviewers for valuable comments.
 This work has been supported by 
the European Community's Seventh
Framework Programme (FP7/2007-2013) under grant agreement No.\ 281043, {}FemtoSpin,
the Swedish Research Council (VR), the Knut and Alice Wallenberg Foundation (Contract No.\ 2015.0060), and the Swedish National  Infrastructure for Computing (SNIC).	 

\end{acknowledgments}

\appendix

\section{Hermiticity of Hamiltonian $\mathcal{H}_{\rm soc}^{\rm ext}$}

The extrinsic spin-orbit Hamiltonian $\mathcal{H}_{\rm soc}^{\rm ext}$, given in Eq.\ (\ref{socext_new}), can indeed be shown to be hermitian, however its individual terms  are not all hermitian. 
Adapting the Einstein summation convention, this Hamiltonian can be written in component form as
\begin{eqnarray}
\!\! \!	\mathcal{H}_{\rm soc}^{\rm ext} = \frac{e}{4m^{2}c^{2}} \! &\Big(& \! i \hbar S_i\partial_t B_i \nonumber \\
       &-& S_i\partial_t B_ir_jp_j+ S_ir_i\partial_t B_jp_j \Big),
	\end{eqnarray}
with $\partial_t \equiv \partial \, / \partial t$.	
To demonstrate that it is hermitian, we take the Hermitian conjugate, and rewrite it in a few steps.
	\begin{widetext}
\begin{eqnarray}
	\Big[\mathcal{H}_{\rm soc}^{\rm ext}\Big]^{\dagger}
	&=&\frac{e}{4m^{2}c^{2}} \Big( -i \hbar S_i\partial_t B_i- S_i\partial_t B_ip_jr_j+ S_i\partial_t B_jp_jr_i \Big) \nonumber\\
	&=& \frac{e}{4m^{2}c^{2}} \Big( -i\hbar S_i\partial_t B_i - S_i\partial_t B_ir_jp_j + S_i\partial_t B_jr_ip_j - S_i\partial_t B_i(p_jr_j) + S_i\partial_t B_j(p_jr_i)
	\Big) \nonumber \\
	&=& \frac{e}{4m^{2}c^{2}} \Big( -i\hbar \bm{S}\cdot\partial_t\bm{B} - (\bm{S}\cdot\partial_t\bm{B})(\bm{r}\cdot\bm{p})+ (\bm{S}\cdot\bm{r})(\partial_t\bm{B}\cdot\bm{p})- (\bm{S}\cdot\partial_t\bm{B})(\bm{p}\cdot\bm{r}) + \bm{S}\cdot\left\{(\partial_t\bm{B}\cdot\bm{p})\bm{r}\right\} \Big) \nonumber \\
	&=& \frac{e}{4m^{2}c^{2}} \Big( -i \hbar \bm{S}\cdot\partial_t\bm{B}-(\bm{S}\cdot\partial_t\bm{B})(\bm{r}\cdot\bm{p})+(\bm{S}\cdot\bm{r})(\partial_t\bm{B}\cdot\bm{p})+{i\hbar}(\bm{S}\cdot\partial_t\bm{B})(\bm{\nabla}\cdot\bm{r})
	- {i\hbar} \bm{S}\cdot\left\{(\partial_t\bm{B}\cdot\bm{\nabla})\bm{r}\right\} \Big) \nonumber\\ 	
	&=& \frac{e}{4m^{2}c^{2}} \Big( -i \hbar \bm{S}\cdot\partial_t\bm{B}- (\bm{S}\cdot\partial_t\bm{B})(\bm{r}\cdot\bm{p})+(\bm{S}\cdot\bm{r})(\partial_t\bm{B}\cdot\bm{p})+3{i\hbar}\bm{S}\cdot\partial_t\bm{B} - {i\hbar}\bm{S}\cdot\partial_t\bm{B} \Big) \nonumber\\ 
	&=& \frac{e}{4m^{2}c^{2}} \Big( i\hbar \bm{S}\cdot\partial_t\bm{B}- (\bm{S}\cdot\partial_t\bm{B})(\bm{r}\cdot\bm{p})+(\bm{S}\cdot\bm{r})(\partial_t\bm{B}\cdot\bm{p}) \Big)
         =	\mathcal{H}_{\rm soc}^{\rm ext} .
\end{eqnarray}
For the individual terms of the Hamiltonian it is straightforward to show their hermitian or non-hermitian character:
\begin{eqnarray}
	\mathcal{H}_{\rm soc}^{\rm ext} =
%
	&=& \underbrace{\frac{ie\hbar}{4m^{2}c^{2}}S_i\partial_t B_i}_{\rm anti-hermitian}-\underbrace{\frac{e}{4m^{2}c^{2}}\sum_{i\neq j}S_i\partial_t B_ir_jp_j}_{\rm non-hermitian}+\underbrace{\frac{e}{4m^{2}c^{2}}\sum_{i \neq j} S_ir_i\partial_t B_jp_j}_{\rm hermitian} .
\end{eqnarray}
As noted before all three terms of the hermitian Hamiltonian contribute to the spin relaxation process. 
\section{From LLG to LL equations of motion}
We found that the generalized LLG equation of spin dynamics can be written in the form [see Eq.\ (\ref{LLG-A})]
\begin{eqnarray}
	\! \! \! \frac{\partial\bm{M}}{\partial t} \! = -\gamma\bm{M}\times\bm{B}_{\rm eff}+\bm{M}\times\Big[\textrm{A}\cdot\frac{\partial\bm{M}}{\partial t}\Big] \, .
\end{eqnarray}
As discussed earlier, using the tensor decomposition, one can also write 
\begin{eqnarray}
	\frac{\partial\bm{M}}{\partial t}=-\gamma\bm{M}\times\bm{B}_{\rm eff}+\alpha\,\bm{M}\times\frac{\partial \bm{M}}{\partial t}+\bm{M}\times\Big[\mathbb{I}\cdot\frac{\partial \bm{M}}{\partial t}\Big]+\bm{M}\times\Big[\bm{D}\times\frac{\partial \bm{M}}{\partial t}\Big]\,.
\end{eqnarray}
The Dzyaloshinskii-Moriya-like damping terms can be expanded, using $\bm{a}\times(\bm{b}\times\bm{c})=\bm{b}(\bm{a}\cdot\bm{c})-\bm{c}(\bm{a}\cdot\bm{b})$,  to give $\bm{M}\times\Big[\bm{D}\times\frac{\partial \bm{M}}{\partial t}\Big]=-\frac{\partial \bm{M}}{\partial t}(\bm{M}\cdot\bm{D})$. Since the magnetization length is conserved we therefore have $\bm{M}\cdot\partial\bm{M}/\partial t =0$.
Defining $(1+\bm{M}\cdot\bm{D})=\Psi$, the LLG equation of spin motion reduces to
\begin{eqnarray}
\Psi\frac{\partial\bm{M}}{\partial t}=-\gamma\bm{M}\times\bm{B}_{\rm eff}+\alpha\,\bm{M}\times\frac{\partial \bm{M}}{\partial t}+\bm{M}\times\Big[\mathbb{I}\cdot\frac{\partial \bm{M}}{\partial t}\Big].
\label{modified_llg}
\end{eqnarray}
Note that $\Psi$ is both a magnetization and Dzyaloshinskii-Moriya vector dependent quantity. Next, we have to calculate the second and third terms on the right-hand side of Eq.\ (\ref{modified_llg}). 
Taking a cross product with $\bm{M}$ on both sides of the last equation gives
\begin{eqnarray}
	\Psi\bm{M}\times \frac{\partial\bm{M}}{\partial t} &=&  -\gamma\bm{M}\times(\bm{M}\times\bm{B}_{\rm eff})+\alpha\,\bm{M}\times\Big(\bm{M}\times\frac{\partial \bm{M}}{\partial t}\Big)+\bm{M}\times\Big(\bm{M}\times\Big[\mathbb{I}\cdot\frac{\partial \bm{M}}{\partial t}\Big]\Big)\nonumber\\
	&=& -\gamma\bm{M}\times(\bm{M}\times\bm{B}_{\rm eff})-\alpha\,M^2\frac{\partial\bm{M}}{\partial t}-M^2\Big[\mathbb{I}\cdot\frac{\partial \bm{M}}{\partial t}\Big]+\bm{M}\Big(\bm{M}\cdot\Big[\mathbb{I}\cdot\frac{\partial \bm{M}}{\partial t}\Big]\Big).
\end{eqnarray} 
Similarly, to evaluate the last term of Eq.\ (\ref{modified_llg}), we take the dot product with the symmetric part of the tensor, followed by a cross product with the magnetization,
\begin{eqnarray}
	\Psi\bm{M}\times\Big[\mathbb{I}\cdot\frac{\partial\bm{M}}{\partial t}\Big]=-\gamma\bm{M}\times\Big[\mathbb{I}\cdot(\bm{M}\times\bm{B}_{\rm eff})\Big]+\alpha\,\bm{M}\times\Big[\mathbb{I}\cdot\Big(\bm{M}\times\frac{\partial \bm{M}}{\partial t}\Big)\Big]+\bm{M}\times\Big(\mathbb{I}\cdot\Big\{\bm{M}\times\Big[\mathbb{I}\cdot\frac{\partial \bm{M}}{\partial t}\Big]\Big\}\Big)\label{mo_modified_llg} .
\end{eqnarray}
At this point we already observe that the first term on the right hand side has adopted a form of the LL damping but with a tensor. The second and third terms are  treated in the following. The second term can be written in component form as 
\begin{eqnarray}
\alpha\,\bm{M}\times\Big[\mathbb{I}\cdot\Big(\bm{M}\times\frac{\partial \bm{M}}{\partial t}\Big)\Big]
=\alpha M_l\mathbb{I}_{mk} M_i\frac{\partial M_j}{\partial t}\varepsilon_{ijk}\varepsilon_{lmn}\hat{\bm{e}}_n .
\end{eqnarray}
We use the following relation for the product of two anti-symmetric Levi-Civita tensors 
\begin{eqnarray}
	\varepsilon_{ijk}\varepsilon_{lmn} = \delta_{il}(\delta_{jm}\delta_{kn}-\delta_{jn}\delta_{km})-\delta_{im}(\delta_{jl}\delta_{kn}-\delta_{jn}\delta_{kl})+\delta_{in}(\delta_{jl}\delta_{km}-\delta_{jm}\delta_{kl}),
	\label{levi}
\end{eqnarray}
 and, defining the trace of the symmetric tensor ${\rm Tr}(\mathbb{I}) =\mathfrak{t} $, a little bit of tensor algebra results in
\begin{eqnarray}
\alpha\,\bm{M}\times\Big[\mathbb{I}\cdot\Big(\bm{M}\times\frac{\partial \bm{M}}{\partial t}\Big)\Big]=\alpha M^2\Big(\mathbb{I}\cdot\frac{\partial \bm{M}}{\partial t}\Big)-\alpha \mathfrak{t} \,M^2\frac{\partial \bm{M}}{\partial t}+\alpha\Big(\bm{M}\cdot\mathbb{I}\cdot\bm{M}\Big)\frac{\partial \bm{M}}{\partial t}-\alpha\bm{M}\Big[\bm{M}\cdot\Big(\mathbb{I}\cdot\frac{\partial \bm{M}}{\partial t} 
\Big)\Big] .
\end{eqnarray}
%
Now we proceed to calculate the last part of Eq.\ (\ref{mo_modified_llg}); the components of this term are given by 
\begin{eqnarray}
	\bm{M}\times\Big(\mathbb{I}\cdot\Big\{\bm{M}\times\Big[\mathbb{I}\cdot\frac{\partial \bm{M}}{\partial t}\Big]\Big\}\Big) &=& M_m\mathbb{I}_{nl} M_k\mathbb{I}_{ij}\frac{\partial M_j}{\partial t}\varepsilon_{kil}\varepsilon_{mno}\hat{\bm{e}}_{o}	 .
\end{eqnarray}
Using once again  the relation in Eq.\ (\ref{levi}) and expanding in different components we find
\begin{eqnarray}
	\bm{M}\times\Big(\mathbb{I}\cdot\Big\{\bm{M}\times\Big[\mathbb{I}\cdot\frac{\partial \bm{M}}{\partial t}\Big]\Big\}\Big) 
	&=& 
	\Big[(\bm{M}\cdot\mathbb{I}\cdot\bm{M})-\mathfrak{t} M^2\Big]\Big(\mathbb{I}\cdot\frac{\partial\bm{M}}{\partial t}\Big)+\Big( \mathfrak{t} \bm{M}-\bm{M}\cdot\mathbb{I}\Big)\Big[\bm{M}\cdot\Big(\mathbb{I}\cdot\frac{\partial\bm{M}}{\partial t}\Big)\Big]\nonumber\\
	&& ~ +(\bm{M}\cdot\bm{M})\Big[\mathbb{I}\cdot\Big(\mathbb{I}\cdot\frac{\partial\bm{M}}{\partial t}\Big)\Big]-\bm{M}\Big[\bm{M}\cdot\left\{\mathbb{I}\cdot\Big(\mathbb{I}\cdot\frac{\partial\bm{M}}{\partial t}\Big)\right\}\Big] .
\end{eqnarray}

Now we have the necessary terms to formulate the LL equation of motion. 
Taking these together, the LLG dynamics of Eq.\ (\ref{modified_llg}) can be written as
\begin{eqnarray}
		\Psi^2\frac{\partial\bm{M}}{\partial t}
	&=& -\gamma\Psi\bm{M}\times\bm{B}_{\rm eff}-\gamma\bm{M}\times\Big[\left(\alpha\mathbb{1}+\mathbb{I}\right)\cdot(\bm{M}\times\bm{B}_{\rm eff})\Big]-\mathbb{G}\cdot\frac{\partial\bm{M}}{\partial t} ,
\end{eqnarray}
with the general tensorial form of $\mathbb{G}$ which is given by
\begin{eqnarray*}
	\mathbb{G}&=& \alpha^2\,M^2\mathbb{1}-\Big[\left(\bm{M}\cdot\mathbb{I}\cdot\bm{M}\right)- \mathfrak{t} M^2\Big]\left(\alpha\mathbb{1}+\mathbb{I}\right)-	
		\Big( \mathfrak{t} \bm{M}-\bm{M}\cdot\mathbb{I}\Big)\bm{M}\cdot\mathbb{I}-
	M^2\mathbb{I}^2+\bm{M}\Big(\bm{M}\cdot\mathbb{I}^2\Big).
\end{eqnarray*}
 Using $\bm{B}= \mu_0 (\bm{H} + \bm{M})$, the transformation from the  LLG to the LL equation results in the form
\begin{eqnarray}
	\Big(\Psi^2\mathbb{1}+\mathbb{G}\Big)\cdot\frac{\partial\bm{M}}{\partial t}=-\gamma_0\Psi\bm{M}\times\bm{H}_{\rm eff}-\gamma_0\bm{M}\times\Big[\left(\alpha\mathbb{1}+\mathbb{I}\right)\cdot(\bm{M}\times\bm{H}_{\rm eff})\Big].
\end{eqnarray}
As mentioned before, in general the Landau-Lifshitz damping cannot be described by a scalar. We find that in the damping term the effect of the anisotropic Ising-like damping is present, while the influence of the Dzyaloshinskii-Moriya-like damping is accounted for through the renormalizing quantity $\Psi$.
\vspace*{0cm}



\end{widetext}
\vspace*{2cm}
\newpage
$\qquad\qquad\qquad\qquad\qquad\qquad\qquad\qquad\qquad\qquad\qquad\qquad$

\section{Expressions for matrix elements}
\label{ab-initio}
We provide here suitable expressions for {\it ab initio} calculations of the matrix elements 
$\langle{r}_{i} {p}_{j}\rangle$. We consider thereto the Bloch states 
$ \vert \nu \bm{k}\rangle$ in a crystal to calculate the expectation value
\begin{eqnarray}
\langle {r}_{i}{p}_{j}\rangle = \sum_{\nu,\nu^{\prime}, \bm{k}}\langle \nu\bm{k}\vert {r}_{i}\vert \nu^{\prime}\bm{k}\rangle\langle \nu^{\prime}\bm{k}\vert {p}_{j}\vert \nu \bm{k}\rangle f(E_{\nu \bm{k}} ),
\label{expect}
\end{eqnarray}
where $f(E_{\nu \bm{k}})$ is the Fermi-Dirac function. 
%
The momentum and position operators are connected  through the Ehrenfest theorem,
$\bm{p}=\frac{im}{\hbar}\left[ \mathcal{H},\bm{r}\right] $,
which we employ to obtain matrix elements of the position operator 
\begin{eqnarray}
\langle \nu^{\prime}\bm{k}\vert\bm{r}\vert \nu\bm{k}\rangle = -\frac{i\hbar}{m}\frac{\langle \nu^{\prime}\bm{k}\vert\bm{p}\vert \nu\bm{k}\rangle}{\left(E_{\nu^{\prime}\bm{k}}-E_{\nu \bm{k}}\right)}.
\end{eqnarray}
Substitution in equation (\ref{expect}) gives
\begin{eqnarray}
\langle {r}_{i} {p}_{j}\rangle &=& -\frac{i\hbar}{m}\sum_{\nu,\nu^{\prime},\bm{k}}f (E_{\nu \bm{k} } )\frac{p_{\nu\nu^{\prime}}^{i} p_{\nu^{\prime}\nu}^{j}}{ E_{\nu\bm{k}}-E_{\nu^{\prime}\bm{k}} } \nonumber \\
&=& -\frac{i\hbar}{2m}\sum_{\nu, \nu^{\prime} ,\bm{k} }\frac{f(E_{\nu \bm{k}}) -f (E_{\nu^{\prime}\bm{k}})}{ E_{\nu\bm{k}}-E_{\nu^{\prime}\bm{k}} } p_{\nu\nu^{\prime}}^{i}p_{\nu^{\prime}\nu}^{j} .
\label{eq:full-rp}
\end{eqnarray}
The double sum over quantum numbers can be further rewritten by separating according to interband matrix elements  ($\nu \neq \nu^{\prime}$) and intraband matrix elements ($\nu = \nu^{\prime}$). The latter part becomes 
\begin{equation}
\langle {r}_{i} {p}_{j} \rangle = -\frac{1}{2}\frac{i\hbar}{m}\sum_{\nu,\bm{k}} \left( \frac{\partial f}{\partial E}\right)_{\! E_{\nu\bm{k}}} \! \! p_{\nu \nu}^{i}  p_{\nu\nu}^{j},
\label{eq:intraband2}
\end{equation}
which can be reformulated using $p_{\nu \nu}^i = \frac{m}{\hbar} \left( {\partial E_{\nu \bm{k}}} / {\partial \bm{k}} \right)_i$ to give expression (\ref{eq:intraband}).
The expressions (\ref{eq:full-rp}) and (\ref{eq:intraband2}) are similar to Kubo linear-response expressions for elements of the conductivity tensor  and  are suitable for first-principles calculations within a DFT framework.

\bibliographystyle{apsrev4-1}

\begin{thebibliography}{85}
\bibitem [{not()}]{note1}%
  \BibitemOpen
  \href@noop {} {}\bibinfo {note} {Note that when the momentum operator and a
  (vector) function are enclosed in round brackets the momentum operator acts
  only on this function.}\BibitemShut {Stop}%
\end{thebibliography}
%

\end{document}